\documentclass[12pt,preprint]{aastex}

\shorttitle{T Pyx}
\shortauthors{Godon et al.}

\begin{document}

\title{
Hubble Space Telescope Far Ultraviolet Spectroscopy of the 
Recurrent Nova T Pyxidis}

\author{
Patrick Godon, Edward M. Sion,             
} 
\affil{Astronomy \& Astrophysics, Villanova University, \\ 
Villanova, PA 19085, USA \\ 
\email{patrick.godon@villanova.edu; edward.sion@villanova.edu} 
\* 
}

\author{
Sumner Starrfield 
} 
\affil{
School of Earth and Space Exploration, Arizona State University, \\ 
Tempe, AZ 85287, USA    \\
\email{sumner.starrfield@asu.edu} 
\* 
}

\author{
Mario Livio, Robert E. Williams, 
} 
\affil{
Space Telescope Science Institute           \\ 
Baltimore, MD 21218, USA   
\email{mlivio@stsci.edu; wms@stsci.edu} 
\* 
}

\author{ 
Charles E. Woodward 
}
\affil{  
Minnesota Institute for Astrophysics, University of Minnesota \\
Minneapolis, MN 55455, USA
\email{chelsea@astro.umn.edu} 
\*
} 

\author{ 
Paul Kuin          
}
\affil{  
Mullard Space Science Laboratory, University College London  \\
Holmbury St Mary, Dorking, Surrey RH5 6NT, UK 
\email{n.kuin@ucl.ac.uk}  
\*
} 

\author{
Kim L. Page
}
\affil{ 
Department of Physics \& Astronomy, University of Leicester \\
Leicester, LE1 7RH, UK
\email{klp5@leicester.ac.uk}  
\* 
} 

\begin{abstract} 

With six recorded nova outbursts, the prototypical recurrent nova T Pyxidis  
is the ideal cataclysmic variable system to assess the net change of
the white dwarf mass within a nova cycle. 
Recent estimates of the mass ejected in the 2011 outburst ranged from 
a few $\sim 10^{-5}M_{\odot}$ to $3.3 \times 10^{-4}M_{\odot}$, 
and assuming a mass accretion rate of $10^{-8}-10^{-7}M_{\odot}$yr$^{-1}$ 
for 44yrs, it has been concluded that the white dwaf  in T Pyx is actually 
losing mass.  
Using NLTE disk modeling spectra to fit our recently obtained 
Hubble Space Telescope (HST) COS \& STIS 
spectra, we find a mass accretion rate of up to two orders of magnitude
larger than previously estimated.    
Our larger mass accretion rate is due mainly to the newly derived   
distance of T Pyx (4.8kpc, larger than the previous 3.5kpc estimate), 
our derived reddening of E(B-V)=0.35
(based on combined IUE and GALEX spectra) and NLTE disk modeling
(compared to black body and raw flux estimates in earlier works). 
We find that for most values of the reddening ($0.25 \le E(B-V) \le 0.50$) 
and white dwaf  mass ($0.70 M_{\odot} \le M_{wd} \le 1.35M_{\odot}$) 
the accreted mass is larger than the ejected mass. 
Only for a low reddening ($\sim 0.25$ and smaller)  {\it combined with} a 
large white dwaf mass ($0.9 M_{\odot}$ and larger) is the ejected mass
larger than the accreted one. However, the best results are 
obtained for a larger value of the reddening.    

\end{abstract} 

\section{INTRODUCTION} 

Cataclysmic variables (CVs) are binaries in which 
a white dwarf (WD; the primary) accretes hydrogen-rich material 
from a Roche-lobe filling secondary star. 
When sufficient matter is accreted, it undergoes a thermonuclear runaway 
(TNR):  the classical nova eruption.  
The TNR ignition starts at the base of the accreted shell when the pressure
and temperature are high enough for the 
CNO cycle burning of hydrogen \citep{pac65,sta72}.  
CVs with recorded multiple nova outbursts 
are classified as recurrent novae (RNe).  
In RNe, a large mass WD and a high mass accretion rate are believed to lead
to the short recurrence time between nova outbursts. 

T Pyxidis, the prototypical RN,  
is known to have erupted \citep{web87,sch13} 
in 1890, 1902, 1920, 1944 and 1967   
- roughly every $\sim$20 yr. It erupted again in April 2011 
\citep{waa11}, after a lapse of 44 years. 
The mass accreted between eruptions and the 
mass ejected during eruptions determine the net change in the 
WD mass. If the WD mass increases, then eventually it
could  reach the Chandrasekhar limit and explode as a type 
Ia supernova (SN Ia; \citet{whe73,ibe84}).
Due to its relatively frequent       outbursts, 
T Pyx offers a promising possiblity             
of assessing the WD mass change within a nova cycle. 
Recent works \citep{sel08,pat13,nel12} 
have claimed that the WD in T Pyx is {\it losing} 
mass and will never become a SN Ia. In this Letter, we present
a synthetic spectroscopic analysis of the first far 
ultraviolet (FUV; 900\AA - 1700\AA ) spectra of T Pyx obtained with Hubble. 
Our objective is to estimate the rate of mass accretion following its July 2011
outburst and compare it with estimates of the mass accretion rate in the quiescent interval 
preceding the July 2011 outburst.

\section{SPECTRAL ANALYSIS} 

The most direct and accurate way to
determine the mass accretion rate of a CV          
accreting at a high mass accretion rate, is to carry out
an analysis and modeling of its ultraviolet 
(UV; $\sim$900\AA - 3200\AA ) spectrum. 
This is because at a high mass accretion rate ($\sim 10^{-8}M_{\sun}$yr$^{-1}$),
an accretion disk around a WD emits more than 50\% of its luminosity in the
UV band (900\AA\ - 3200\AA\ ) and contributes much less to
the optical. A significant fraction of the remaining energy is
emitted in the extreme UV (EUV; $\lambda \sim$ 100\AA - 900\AA ), 
and that fraction increases with increasing mass accretion rate.

\subsection{Hubble STIS + COS Spectroscopy of T Pyx}   

T Pyx was observed in the UV ($\sim$1150\AA - 3200\AA )             
first with the {\it International Ultraviolet Expolorer} 
(IUE). 
We co-added and combined 29 IUE spectra obtained with the Short
Wavelength Prime (SWP; 1150\AA\ - 2000\AA\ ) camera and 11 IUE spectra 
obtained with the Long Wavelength Prime (LWP; 1850\AA\ - 3200\AA\ ) camera. 
These spectra, obtained between 1986 and 1996,
exhibit the same continuum flux level. 
The IUE spectra show that the UV continuum of T Pyx remained nearly 
constant in slope and intensity, without any indication of long term trends,  
while the  emission lines exhibit  changes in both intensity and 
detectability as already mentioned by \citet{gil07}. 

T Pyx was observed with the {\it Galaxy Evolution Explorer} 
(GALEX) at the end of 2005.
The GALEX spectrum is composed of two spectra, one 
in the FUV band (1350\AA\ -1800\AA ) and one 
in the near UV (NUV) band (1800\AA\ -3000\AA ). 
In spite of detector edge problems, 
the GALEX and IUE spectra match remarkably well
in those regions away from the edges,
in agreement with the suggestion \citep{gil07}   
that the UV continuum remains constant over the years. 
This is an indication that {\it the mass accretion rate itself is constant},
since the UV emission comes mainly from the accretion disk.  
Consequently, the steady decline of the optical magnitude of T Pyx, 
observed since 1890, does not indicate                  a decrease in 
the mass accretion rate. 

FUV (1150\AA - 2250\AA ) spectra of T Pyx were obtained with the 
HST/{\it Space Telescope Imaging Spectrograph} (STIS) 
during the eruption and the decline to quiescence \citep{sho13}.  
We further followed T Pyx into quiescence, obtaining
HST/STIS spectra ($\sim$1150\AA\ - 1700\AA\ )
together with HST/Cosmic Origins Spectrograph (COS) spectra 
($\sim $900\AA\ - 1200\AA\ ) in March 2012, December 2012 and July 2013. 
While the March 2012 spectrum still exhibits strong emission lines
and a flux larger than the pre-outburst flux level, the December 2012
and July 2013 spectra have weak emission lines and have reached 
the pre-outburst flux level of the IUE and GALEX spectra. 
Except for small differences in emission lines, 
the December 2012 and July 2013 HST spectra are identical,
and show that T Pyx 
has now returned  precisely to its pre-outburst value (i.e. IUE
flux level). 
This indicates  that the mass accretion rate has      
returned to its pre-outburst value. 

Since the December 2012 and July 2013 HST spectra are almost identical,
we co-added them to improve the signal-to-noise (S/N) 
in preparation for the spectral model fit.  
In Fig.1 we show the quiescent co-added 
IUE, GALEX and co-added HST spectra after removal of
the noisy detector edges. The spectra, though 
acquired with different telescopes and instruments, match accurately.  
The minimum near 2175\AA\ 
is due to interstellar extinction. Since the GALEX spectrum
is the most reliable 
in that wavelength region, we combine it        
with the IUE co-added spectrum to determine E(B-V).
We then deredden the combined spectrum for different values of E(B-V). 
The E(B-V) value for which the 2175\AA\ feature disappears,  
E(B-V)=0.35 (see Fig.2), is taken as the E(B-V) value 
towards T Pyx, in agreeement with \citet{bru81}.
We use this value to deredden the IUE, 
GALEX and HST spectra, but we also consider the effects of different
reddening values on our results.

We combined the HST (STIS+COS), IUE and GALEX spectra, excluding 
the noisy portions,
and obtained a spectrum from 
$\sim 900$\AA\ to $\sim 3200$\AA . While we modeled the entire
spectrum, we never expected the model to fit in the NUV range
($\sim$2000\AA - 3000\AA ), because
we only modeled the hottest component of the system, the accretion disk,
which mainly contributes to the FUV ($\sim$900\AA - 1700\AA ) 
and extreme UV ($\sim$100\AA - 900\AA ).
The shorter wavelengths of the spectrum covered by COS are crucial in
the determination of the mass accretion rate as this is where the spectrum
is expected to peak if the accretion rate is large.  

\subsection{Modeling} 
The disk is the standard disk model \citep{sha73,pri81}, and for 
a given mass accretion rate and a given WD mass
it is divided in  $N$ rings with temperature   
$T_i, (i=1,2,.. N)$ located at $r=r_i, (i=1,2,..N)$, 
where $T_i=T_{eff}(r_i)$, and $T_{eff}(r)$ is the radial temperature
profile of the standard disk model (e.g. \citet{pri81}). 
Each ring is modeled using the stellar
atmosphere code TLUSTY, 
then a synthetic spectrum is computed with 
the code SYNSPEC for each ring, and the spectra are finally combined together 
with DISKSYN taking into account Keplerian broadening, 
inclination and limb darkening \citep{hub88,hub95,wad98}. 
For rings with $T>30,000$K TLUSTY and SYNSPEC are run with basic NLTE
options.  

To emphasize the importance of modeling the disk in the UV, 
we use our disk modeling to
check what fraction of the disk luminosity is emitted in the
UV band (900\AA $< \lambda_{UV} <$ 3200 \AA ). 
For a cannonical WD mass of $0.8M_{\odot}$, 
at a mass accretion rate of $10^{-9.5}M_{\odot}$yr$^{-1}$ the disk emits 
58\% of its luminosity in the UV band ($\lambda_{UV}$),
and most of the remaining luminosity ($\sim$40\%) is emitted
at longer wavelengths. 
As the mass accretion rate increases, so does the temperature in the disk. 
At $\dot{M}=10^{-9}$ the disk emits 66\% of its luminosity in the UV,
about 30\% at longer wavelengths and only a few percent in the EUV 
($\lambda < 900$\AA ). 
As the mass accretion rate keeps increasing, the disk starts emitting
more flux in the EUV and less flux in the UV and optical, as the
Planckian peak shifts to shorter and shorter wavelengths.   
At $\dot{M}=10^{-8}M_{\odot}$yr$^{-1}$, the disk emits $\sim$50\% in the
UV with about 25\% emitted at shorter wavelengths and 25\% emitted at longer
wavelengths. The amount emitted in the visual band is just a fraction of
the 25\% and is always much less than the amount emitted in the UV band.   
This justifies the modeling of accretion disk in the UV band. 
The main source of the optical luminosity  is not from the disk, but most likely
arises from the contribution from the hot spot, the secondary, and
possibly from some of the nebular material.  

From theoretical predictions, to reproduce an outburst every 
20yrs or so, the WD mass in T Pyx is expected to be very large
(almost near-Chandrasekhar). On the other hand, the derived
mass ratio $q=0.2$ in the system together with an {\it anticipated} 
secondary mass of $M_2=0.14M_{\odot}$ imply a WD mass of 
only $M_{wd}=0.7M_{\odot}$ \citep{uth10}. 
However, the secondary has never been 
spectroscopically  detected to verify the assumption $M_2=0.14M_{\odot}$.   
Consequently, due to the uncertainty in the WD mass, 
we generated disk models for an accreting WD with a near Chandrasekhar mass 
of $1.35M_{\odot}$ as expected from the observed short 
recurrence time of T Pyx \citep{sta85,web87,sch10},
and for a lower mass WD, $0.70M_{\odot}$,  
as inferred by \citet{uth10}.
The latter low mass, if confirmed,
raises serious problems for the theory of recurrent shell flashes.
We varied the mass accretion rate, $\dot{M}$, 
from $\sim 10^{-9} M_{\odot}$yr$^{-1}$ to $\sim 10^{-6} M_{\odot}$yr$^{-1}$. 
The inclination ($i$) of the system is low with a lower limit of 
about $i=10^{\circ}$ \citep{uth10} 
and an upper limit of  $i=30^{\circ}$ \citep{web87}. 
Since we model only the continuum, 
the effect of the inclination on the results is very small, as long as the inclination is low. 
For a given WD mass (and therefore radius), the fitting of the observed 
spectrum with a theoretical accretion disk spectrum is carried out by 
scaling the theoretical flux to the distance of 4.8kpc, 
which has been recently derived using the light echo technique
\citep{sok13}. 

\section{RESULTS} 

We first fit the dereddened spectrum 
assuming E(B-V)=0.35. 
For the $M_{wd}=1.35M_{\odot}$ case, we found a mass accretion rate of 
$\dot{M}= 1.6 \times 10^{-6} M_{\odot}$yr$^{-1}$ for $i=10^{\circ}$
and   
$\dot{M}= 1.9 \times 10^{-6} M_{\odot}$yr$^{-1}$ for $i=30^{\circ}$.
The total mass accreted over 44yr (since the previous
explosion) is $7.0 \times 10^{-5}M_{\odot}$ and $8.4 \times 10^{-5}M_{\odot}$, 
for $i=10^{\circ}$ and $i=30^{\circ}$ respectively.  
For the $M_{wd}=0.70M_{\odot}$ case, the mass accretion rate needed to
fit the spectrum increases to $\dot{M}=2.2 \times 10^{-6}M_{\odot}$yr$^{-1}$
and $2.7 \times 10^{-6}M_{\odot}$yr$^{-1}$,   
for $i=10^{\circ}$ and $i=30^{\circ}$ respectively;  and the total
accreted mass becomes  
$9.7 \times 10^{-5}M_{\odot}$ and $1.2 \times 10^{-4}M_{\odot}$. 
A synthetic accretion disk model spectral fit 
to the HST/COS spectrum is presented in Fig.3, 
for $M_{wd}= 0.7M_{\odot}$ and an inclination of $i=20^{\circ}$. 
The exact same spectral fit to the 
HST/STIS spectrum and combined GALEX IUE spectrum is presented
in Fig.4.  
The synthetic spectrum is deficient in
flux in the longer wavelength range, $\lambda > 1600$\AA\ , indicating
the possible contribution of a colder component ($T<10,000K$).   

Even though we derived a reddening of 0.35, which is consistent with
the originally derived E(B-V) value from the IUE data in \citet{bru81},  
we nevertheless checked the effect of the reddening on the results
(since different methods and reddening curves for evaluating 
E(B-V) lead to different results (e.g. 
\citet{sho13}; see also \citet{gil07}). 

Based on IUE data alone, \citet{gil07} 
derived a reddening value of  $E(B-V)= 0.20-0.25$. 
We therefore checked our fitting results  
for the value of $E(B-V)= 0.25$, and we found that 
to fit the distance of 4.8kpc one needs a mass accretion rate of 
$9 \times 10^{-7}M_{\odot}$yr$^{-1}$ (i=20deg). However, 
this synthetic spectrum does not fit the slope 
of the observed spectrum, the observed spectrum is too {\it red}. 
To fit the slope of the spectrum one needs a much lower mass
accretion rate, namely $\dot{M}=10^{-8}M_{\odot}$yr$^{-1}$, but
the resultant produced flux level is lower which requires  
a distance to T Pyx of only 1250pc. Similar results are obtained
for this reddening when assuming a 1.35$M_{\odot}$ WD mass. 
 
Next, for a larger reddening value of $E(B-V)=0.5$ \citep{sho13} the
observed spectrum is too {\it blue}, and our models are limited by 
the constraint that TLUSTY does not handle mass accretion rates greater
than $5 \times 10^{-6} M_{\odot}$yr$^{-1}$ for a $0.7M_{\odot}$ WD mass
and  
$1 \times 10^{-6} M_{\odot}$yr$^{-1}$ for a $1.35M_{\odot}$ WD mass. 
We extrapolated linearly our results
by matching flux level; namely, we assume that 
the flux level is a linear function of the
mass accretion rate and we ignore Wien's displacement. 
Since the Planckian peaks move towards shorter wavelengths
as $\dot{M}$ increases, this extrapolation understimates $\dot{M}$, 
and this effect is more pronounced as $\dot{M}$ increases.  
Using this technique, we found that the mass accretion rate needed 
to fit the observed spectrum assuming $E(B-V)=0.5$ for a  
$M_{wd}=0.7 M_{\odot}$ (as  
in \citet{sho13}) is $\dot{M}=3 \times 10^{-5}M_{\odot}$yr$^{-1}$,
implying a total accreted mass of $1.32 \times 10^{-3}M_{\odot}$. 
For the $1.35M_{\odot}$ WD mass case, this linear extrapolation
gives a mass accretion rate of $8 \times 10^{-6}M_{\odot}$yr$^{-1}$
and a total accreted mass of $3.5 \times 10^{-4}M_{\odot}$.

\section{CONCLUSIONS} 

Our results are recapitulated in Fig.7,    
where we draw (solid lines) the mass accreted over 44 years 
as computed from our model fittings, for 3 different values of
the reddening.  We also draw (dotted line) the lower limit 
for the mass accreted derived from integrating the UV flux  
over the wavelengths 900\AA\ to 3200\AA\  
$$ 
\Delta M_{min} = 44 \times \dot{M}_{min} = 
44 \times \int F_{\lambda} d\lambda ~. 
$$    
Since the disk at high mass accretion rate emits only a
fraction of its energy in the UV (and a much smaller fraction
is emitted in the optical), the integrated UV flux is clearly a lower limit
for the mass accretion rate. 
We draw this lower limit for three
different values of the reddening as well. For comparison we draw 
the ejecta mass as derived by radio observations
by \citet{nel12} (denoted with a square, \#1), who assumed  
$M_{wd}=0.7M_{\odot}$ and E(B-V)=0.50. This ejecta  
mass is to be compared with our accreted mass derived from our
disk modeling assuming E(B-V)=0.50. Our results show that
the accreted envelope (for the particular values of the parameters) 
is about 4 times larger than the mass ejected as estimated by 
\citet{nel12}. 
On the graph, we also show the mass of the ejected envelope as estimated
by \citet{pat13}  assuming that energy and momentum are conserved
(dashed line, \#2).  We see that the disk model
fittings imply that more mass is accreted than ejected, except
when assuming a reddening of E(B-V)=0.25 {\it in combination with} 
a WD          mass $M_{wd}> 0.9M_{\odot}$.   However, for this value of
the reddening, the disk model (for all WD masses considered) did
not fit the observed spectrum, indicating that the reddening might
indeed be larger (i.e. 0.35).  

We note that the $\sim 10$\% error in the distance $4.8 \pm 0.5 $kpc 
translates into an error of the order of 20\% in the mass accretion
rate, which does not change our results significantly.  
We also emphasize that 
the difference between our results and previously derived
mass accretion rates is due mainly to the now larger distance 
estimates (4.8kpc vs 3.5kpc) as well as a larger reddening
(0.35 against 0.25). 
 
Our      results are: (i) the determination of 
the long-time-baseline accretion rate in T Pyx, using for the
first time realistic disk models, 
(ii) the demonstration that the 
accretion rate in T Pyx has returned to exactly its pre-outburst UV 
spectral energy distribution. 
This accretion rate, $\sim 10^{-6}M_{\odot}$yr$^{-1}$ (and higher for
increasing values of the reddening), 
is much higher than previously estimated. 

It is possible that T Pyx sustains such a high mass accretion rate
because of irradiation of the secondary star. The irradiation
of the donor star by the extremely hot WD and inner disk will not only
``puff up'' the donor's radius but may also drive a wind off the secondary
\citep{kni00}. Since the orbital period and binary separation of T Pyx 
are smaller than all other known recurrent novae, the effects of irradiation
may be more pronounced.  

Several different authors reported that quasi-static evolutionary sequences of accreting WDs
expand in to red giant structures at accretion rates exceeding 
$10^{-7} M_{\odot}$yr$^{-1}$ 
\citep{pac78,sio79,ibe82}  
but other evolutionary accreting WD          sequences followed 
hydrodynamically through multiple nova outbursts 
(e.g. \citet{pri95,yar05,sta12,ida13})  
reveal thermonuclear 
shell flashes at these high rates.

\noindent 
{\bf Acknowledgements} This research is supported by HST grants
GO-12799.01A and GO-12890.01A, both to Villanova University.
Support for the analysis of the IUE and 
GALEX archival data was provided by the National Aeronautics and Space 
Administration (NASA) under grant number NNX13AF11G issued through the
Office of Astrophysics Data Analysis Program (ADP) to Villanova University. 
P.G. thanks William P. Blair for his kind hospitality at the Henry A.
Rowland Department of Physics and Astronomy at the Johns Hopkins University,
Baltimore, MD. We also thank Steve Shore for his comments 
on the manuscript.

\clearpage

\begin{figure}
\vspace{-8.cm} 
\plotone{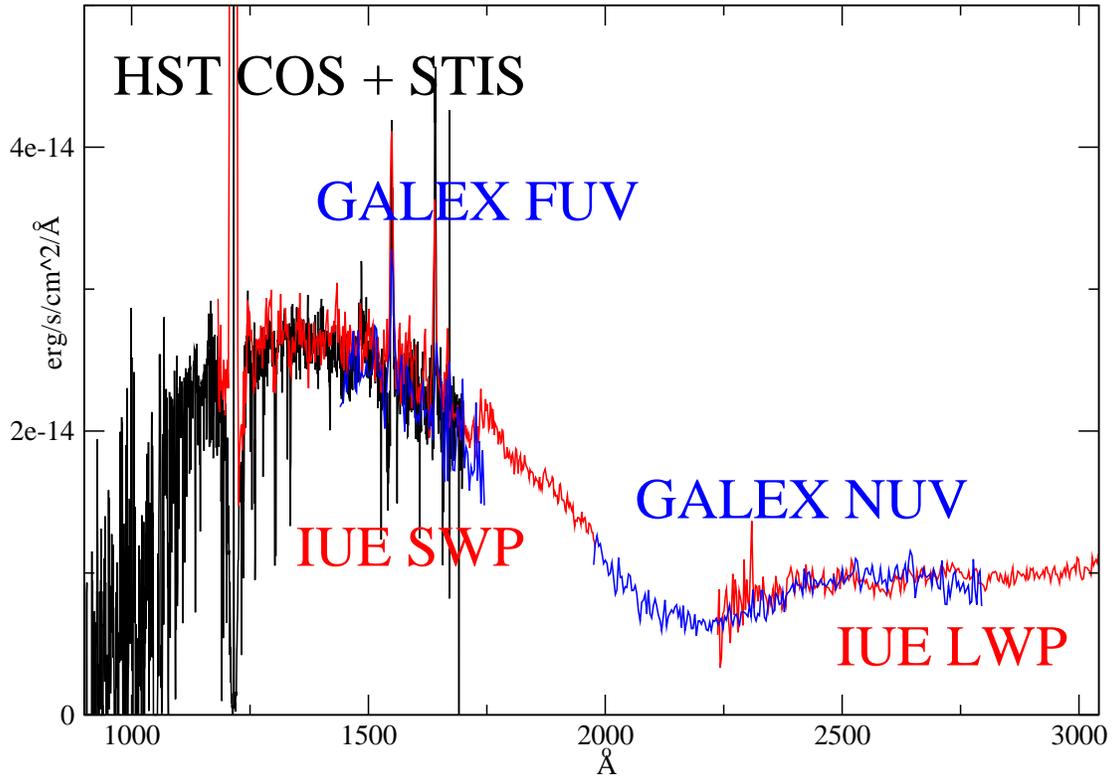}  
\vspace{-1.0cm} 
\caption{
The existing UV spectra of T Pyx before and after its eruption. 
The archival IUE SWP and LWP spectra were obtained 
over a period of 10 years and match both the GALEX and 
HST spectra. The noisy edges of the detectors have been removed
for clarity.  
The UV ($\sim$1000\AA - 3000\AA )
flux of T Pyx has remained fairly constant since
it was first observed with IUE. This indicates that the mass accretion
rate itself remained constant, and after its 2011 eruption it 
returned to its pre-outburst value. 
 }
\end{figure}

\clearpage

\begin{figure}
\vspace{-5.cm} 
\plotone{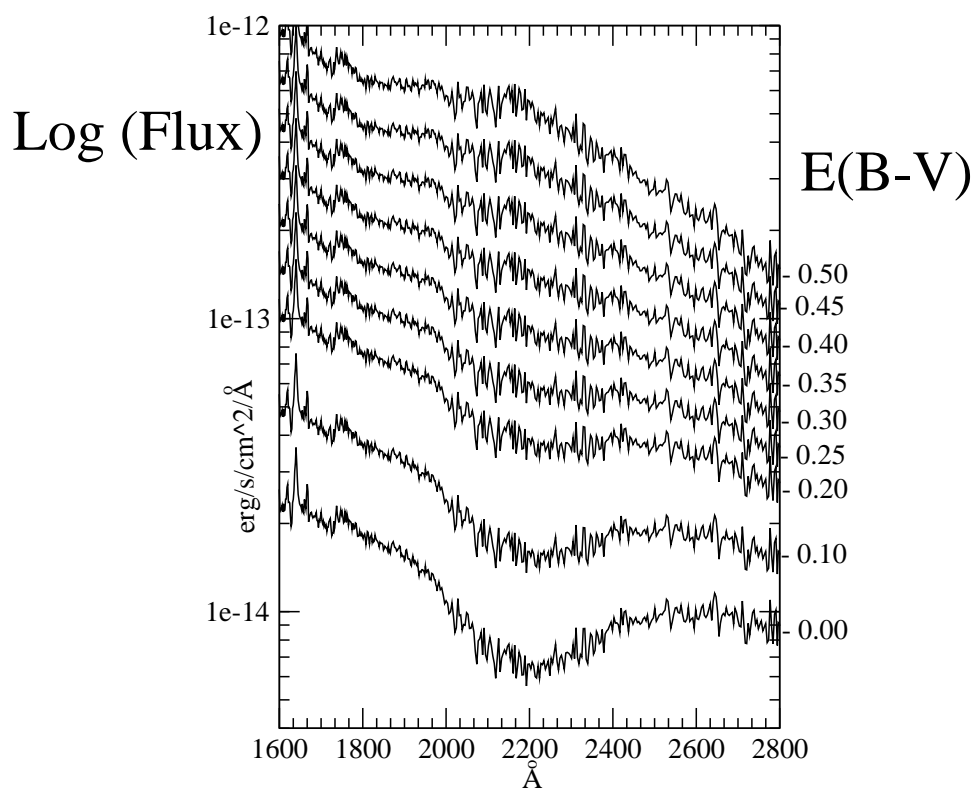} 
\vspace{-1.2cm} 
\caption{The merged GALEX-IUE spectrum (see Fig.1) has been dereddened
for different values of E(B-V) as indicated on the right. The 2175\AA\ 
feature associated with the reddening is clearly seen in absorption 
for low values of E(B-V), and it appears as extra flux (`emission'
like) for large values of E(B-V). We deduce that the
reddening towards T Pyx must be E(B-V)=0.35, the value for which 
the 2175\AA\ region vanishes (i.e. it becomes rather flat and featureless 
on the Log graph).  
 } 
\end{figure}

\clearpage

\begin{figure}
\vspace{-9.cm} 
\plotone{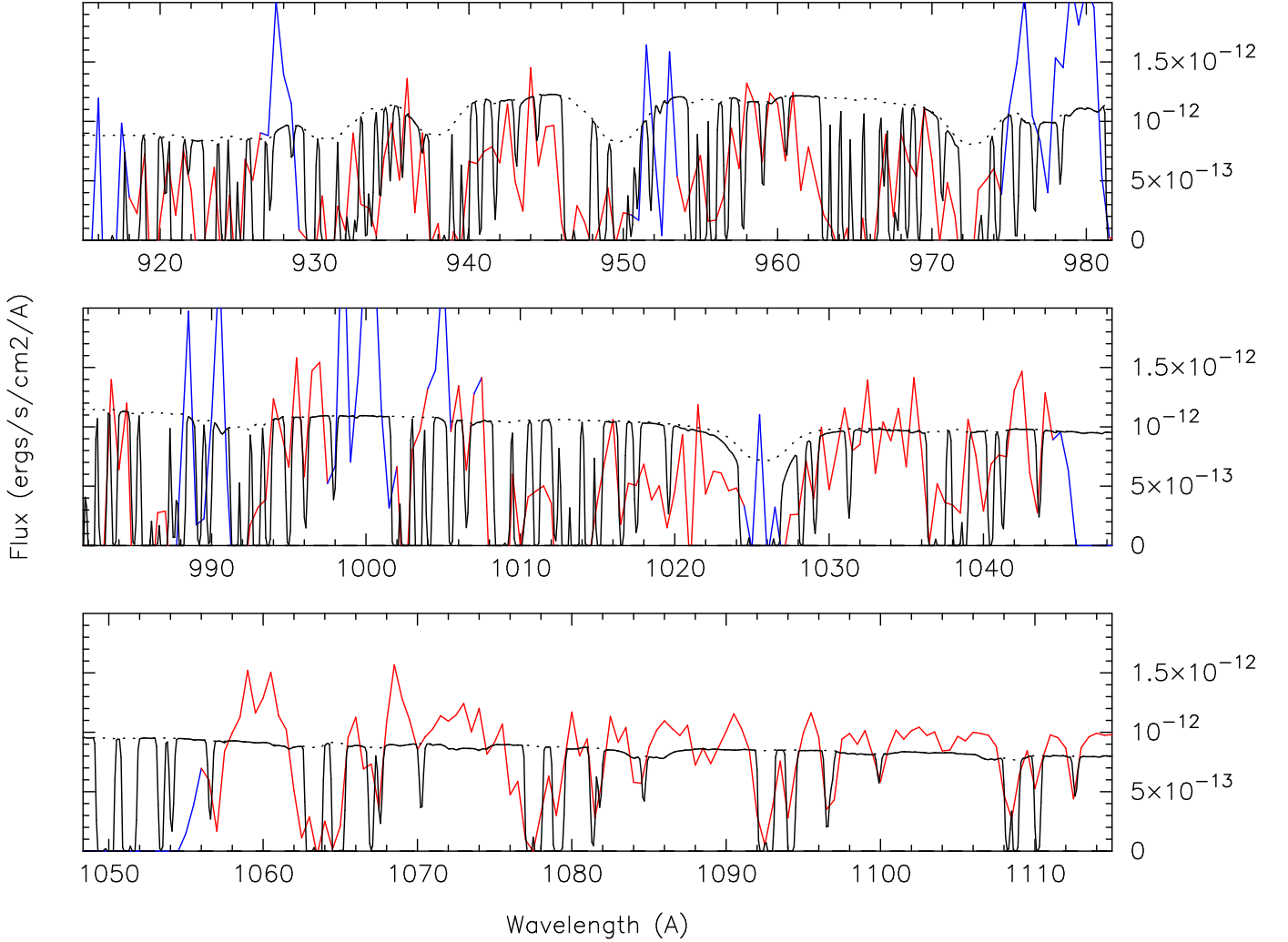}
\vspace{0.cm} 
\caption{The HST COS spectrum (in red) of T Pyx
has been fitted with a disk model (solid black line) 
including an interstellar (ISM) absorption
curtain (the dotted line indicates the disk model without ISM absorption). 
The COS data are rather noisy. The regions marked in blue are either 
due to noise, airglow, or emission lines, and are not modeled. 
The fitting is carried out between the solid black line 
and the solid red line. 
The gap between 1045\AA\ and 1055\AA\ is a detector gap. 
The spectrum has been dereddened assuming E(B-V)=0.35. 
At a distance of 4.8kpc, the flux
in the COS spectrum is accounted for with a large mass accretion rate. 
In this model we have $M_{wd}=0.7 M_{\odot}$, $i=20^{\circ}$, 
$\dot{M} = 2.4 \times 10^{-6} M_{\odot}$yr$^{-1}$. For an accretion period
of 44yr, this gives a total accreted mass of $\sim 10^{-4}M_{\odot}$.     
 } 
\end{figure}

\clearpage 

\begin{figure}
\vspace{-20.cm} 
\plotone{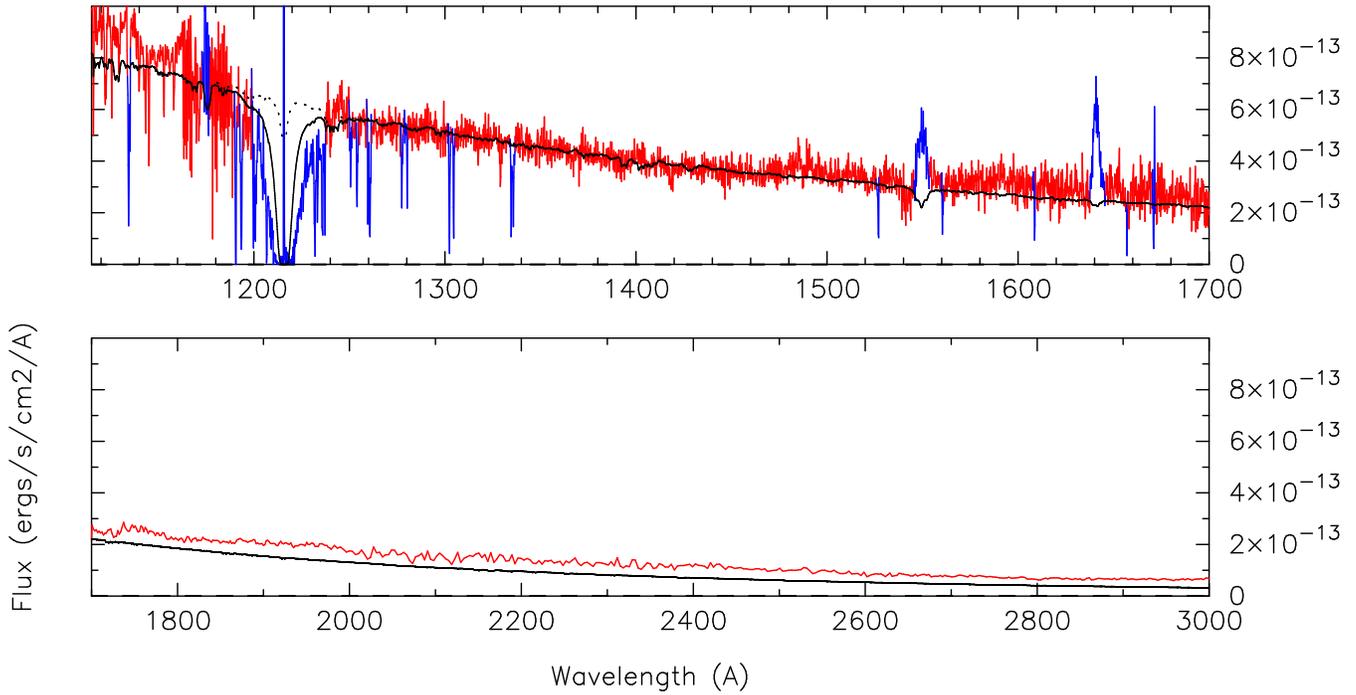} 
\vspace{-3.cm} 
\caption{
The same model fit shown in Fig.1, is shown here 
in the longer wavelengths, covering 
mainly the STIS and IUE range. 
The Ly$\alpha$ region ($\sim$1200\AA\ -1240\AA\ ) 
is affected by the ISM and is therefore not modeled. The sharp
emission and absorption lines are also not included in the modeling.  
From $\sim$1600\AA\ and longward into the IUE LWP range the observed spectrum  
has extra flux possibly due to a colder component in the system.   
 } 
\end{figure}

\clearpage 

\begin{figure} 
\vspace{-8.cm} 
\plotone{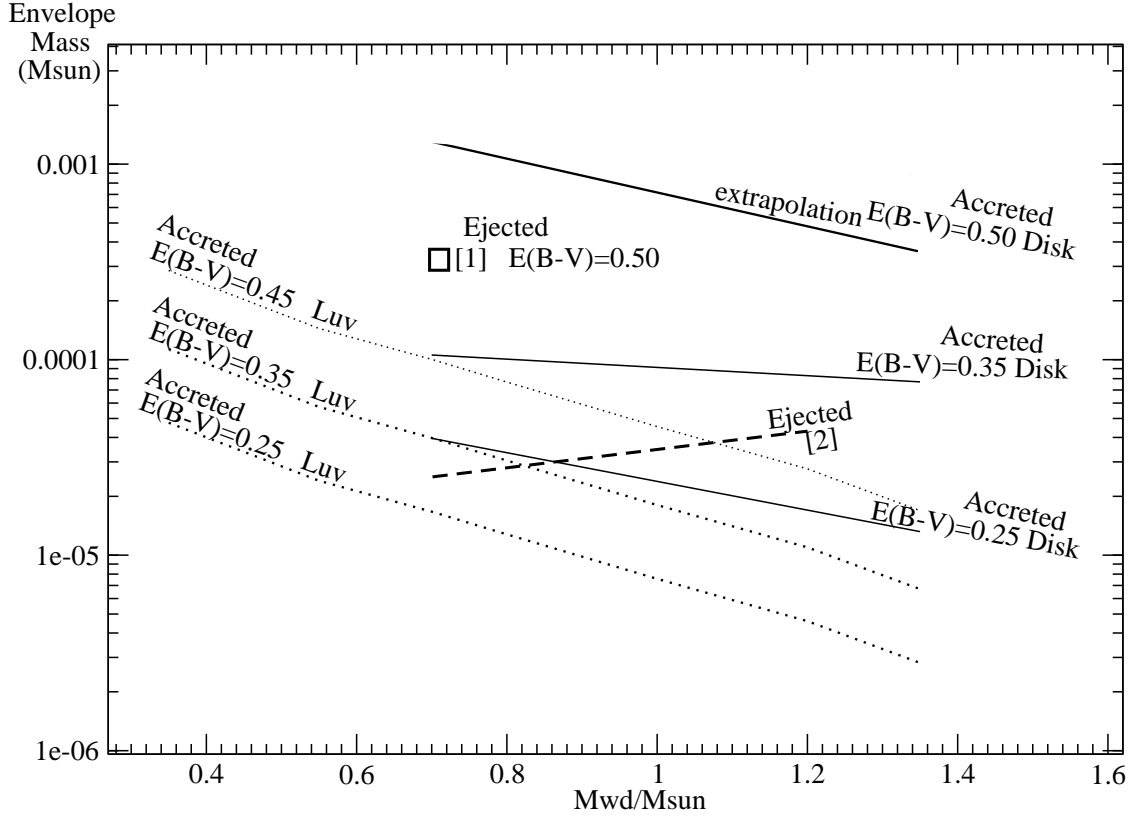}
\caption{
The mass of the accreted envelope and ejected envelope are shown as a function
of the WD          mass for different values of the reddening. 
Our disk model results are drawn with 
a solid line (``Disk''). 
The lower limit of the accreted envelope 
inferred from the UV flux is shown (dotted line, ``Luv''). 
In comparison we show the ejected envelope [1] as estimated by \citet{nel12}  
(square symbol), as well as [2]  
computed in \citet{pat13} (dashed line). The ejected estimate [1] was 
computed assuming E(B-V)=0.5 and $M_{wd}=0.7M_{\odot}$ and is to be 
compared with our result for the same value of E(B-V) and $M_{wd}$. 
The accreted envelope is always larger than
the ejected envelope, except for a value of  E(B-V)=0.25 {\it combined with} 
a WD mass $\sim 0.9M_{\odot}$ and larger. However, the best results
were obtained for a larger value of the reddening $E(B-V) > 0.30 $, 
as the E(B-V)=0.25
value led to either a very bad fit or a distance of only $\sim 1$kpc. 
The accreted envelope for the E(B-V)=0.50 case 
(marked ``extrapolation'') has been
carried out by linearly extrapolating our accretion disk models, 
and is understimate (this is more pronounced for the $1.35M_{\odot}$ 
WD mass case).
} 
\end{figure}

\end{document}